\documentclass[aps,fleqn,pre,showpacs,twocolumn]{revtex4}

\usepackage{amsmath,amssymb,graphicx}

\newcommand{\knni}[1][i]{k^{(i)}_{\mathrm{nn}}}
\newcommand{\knnl}[1][i]{k^{(l)}_{\mathrm{nn}}}

\newcommand{\tknni}[1][i]{\tilde{k}^{(i)}_{\mathrm{nn}}}

\begin{document}

\title{Generating random networks with given degree-degree correlations and degree-dependent clustering}

\author{Andreas~Pusch}
\affiliation{Institut~f\"ur~Festk\"orperphysik,
             Technische~Universit\"at~Darmstadt,
             Hochschulstr.~8, 64289~Darmstadt, Germany}

\author{Sebastian~Weber}
\affiliation{Institut~f\"ur~Festk\"orperphysik,
             Technische~Universit\"at~Darmstadt,
             Hochschulstr.~8, 64289~Darmstadt, Germany}

\author{Markus~Porto}
\affiliation{Institut~f\"ur~Festk\"orperphysik,
             Technische~Universit\"at~Darmstadt,
             Hochschulstr.~8, 64289~Darmstadt, Germany}

\date{\today}

\begin{abstract}
Random networks are widely used to model complex networks and research their properties. In order to get a good approximation of complex networks encountered in various disciplines of science, the ability to tune various statistical properties of random networks is very important.
In this manuscript we present an algorithm which is able to
construct arbitrarily degree-degree correlated networks with adjustable degree-dependent clustering. We verify the algorithm by using empirical networks as input and describe additionally a simple way to fix a degree-dependent clustering function if degree-degree correlations are given.
\end{abstract}


\maketitle

\section{Introduction}
\label{sec:introduction}

Modeling empirical networks as random networks is an important approach in the effort of studying topology and dynamics of complex networks. The first attempts in constructing random networks which exhibit some of the common features regularly found in empirical networks from fields as different as biology, social sciences, and technology have mostly aimed at understanding the origin of scale-free degree distributions (the degree of a vertex being its number of connections) and small average distances among vertices \cite{Barabasi1999,Watts1998}. However, it has been found that there are other important statistical quantities that profoundly influence the structure of complex networks and consequently the dynamics taking place on them. Notably among them are degree-degree correlations of vertices \cite{Vazquez2002, Newman2003a, Newman2003} and the abundance of motifs \cite{Barabasi2003,Milo2002, Itzkovitz2005}. The smallest and probably most important motif in undirected graphs is the triangle. Its abundance is called clustering and several measures have been proposed to quantify it \cite{Serrano2007}. Some refined network growing mechanisms which extend the preferential attachment scheme introduced by \citeauthor{Barabasi1999} to generate  ``scale-free'' graphs \cite{Barabasi1999} that are either correlated or clustered have been proposed \cite{Holme2002,Vazquez2003, Newman2003b,  Satorras2005}. Those algorithms are, however, restricted in the correlation and clustering patterns they are able to produce.

Therefore, some efforts have recently been undertaken to overcome these restrictions. For example, the very successful configuration model (CM) algorithm \cite{Molloy1995,Molloy1998,Catanzaro2005}, capable of
generating random networks with an {\em a priori} given degree distribution, has been extended to include either degree-degree correlations or clustering properties of networks. \citeauthor{Serrano2005} presented an algorithm capable of tuning the degree-dependent clustering coefficient as well as the degree distribution \cite{Serrano2005}. Additionally, they pointed out that clustering and degree-degree correlation are deeply entwined, the latter limiting the former especially for vertices of high degree, in particular for disassortative networks where vertices of high degree are preferentially connected to vertices of low degree and vice versa.
As both properties, clustering and correlations, are very important for the structure of a network and strongly related to each other, it is a natural ansatz to control degree-dependent clustering and the correlation pattern simultaneously to achieve better null models of complex networks.
In this manuscript, we propose an algorithm to construct random networks with given degree-degree correlation structure and degree-dependent clustering. It is organized as follows:
Section II introduces the network clustering and correlation measures used. Section III describes the algorithm to construct degree-degree correlated and clustered networks and verifies our scheme by applying it to empirical networks.
Section IV presents a simple way to create networks with certain correlations and clustering and shows some results of this approach. In Section V we briefly summarize.

\section{Network Correlation and Clustering Measures}
\label{sec:corNetworks}

Two-point degree-degree correlations can statistically be described via a degre-degree correlation function $P(j,k)$ which is the probability that a randomly chosen edge has vertices of degrees $j$ and $k$ at its ends. In the case of uncorrelated networks, the correlation function factorizes into $P^{u}(j,k) = kP(k)jP(j)/\langle k \rangle^{2}$, where $P(k)$ is the degree distribution. Thus it appears natural to define a correlation function $f(j,k)$ as
\begin{equation}
 \label{eq:fjkDef}
f(j,k) = \frac{P(j,k)}{P^{u}(j,k)}.
\end{equation}
Values of $f(j,k)$ different from $1$ signal degree-degree correlations in the underlying network.
A simpler but more coarse-grained manner to quantify degree-degree correlations is the average nearest neighbor function $k_{\mathrm{nn}}(k)$, describing the average degree of neighbors of vertices with degree $k$. It can be calculated from the conditional probability $P(j|k) = P(j,k) \left\langle k \right\rangle / [kP(k)]$ as
\begin{equation}
\label{eq:knnDef}
k_{\mathrm{nn}}(k) = \sum_{j} j P(j|k).
\end{equation}
A network with an (de-)increasing $k_{\mathrm{nn}}(k)$ is called (dis-) assortatively correlated.

Clustering was originally defined by Watts and Strogatz \cite{Watts1998} for the vertex $i$ to be
 \begin{equation}
   \label{eq:clusterDef}
   c_i = \frac{2 T_i}{k_i (k_i - 1)},
 \end{equation}
 where $T_i$ denotes the number of triangles passing through vertex
 $i$. Clearly this measure is a three-point dependent value as the
 number of triangles requires knowledge over three connected vertices
 at the same time. However, it is common use to average the clustering
 coefficients $c_i$ of all vertices with the same degree $k$ together,
 yielding a degree-dependent clustering coefficient
\begin{equation}
c(k) = \frac{1}{k(k-1)P(k)N} \sum_{i \in \Upsilon(k)} 2 T_{i} \, ,
\end{equation}
where $\Upsilon(k)$ denotes the set of vertices with degree $k$.

\citeauthor{Serrano2005} pointed out that the degree-dependent clustering $c(k)$ is restricted by degree-degree correlations and is often found to be a decreasing function of $k$ \cite{Serrano2005}. They calculated an upper limit $\lambda(k)$ of $c(k)$ dependent on the degree-degree correlation function $P(j,k)$. The main reasoning is that an edge cannot be part of more triangles than $\min(k_{i}, k_{j})-1$ with $k_{i}$ and $k_{j}$ being the degrees of the vertices connected by it. This results in a constraint on the number of triangles $T_{i}$ for any vertex $i$,
\begin{equation} 
\label{eq:numRest}
T_{i} \leq \sum_{j} a_{ij} [\min(k_{i}, k_{j})-1] \, .
\end{equation}
Here $a_{ij}$ is the network's adjacency matrix. The upper limit $\lambda(k)$ of the degree-dependent clustering $c(k)$ can than be written as
\begin{equation}
  \label{eq:lambdaDef}
\lambda(k) \equiv 1 - \frac{1}{k-1} \sum_{j=1}^k (k-j) P(j|k) \geq c(k) \, .
\end{equation}
This function is always a decreasing function of $k$ and its slope depends strongly on the average neighbor degree $k_{\mathrm{nn}}(k)$.
This means that degree-dependent clustering $c(k)$ can be written as
\begin{equation}
 \label{eq:ckbylambda}
c(k) = c_{\mathrm{eff}}(k) \lambda(k)
\end{equation}
with $ 0 \leq c_{\mathrm{eff}}(k) \leq 1 \, \forall \, k$. Thus $c_{\mathrm{eff}}(k)$ can be regarded as an effective degree-dependent clustering, once degree-degree correlations are fixed.

In the following, we describe an algorithm that is able to control the two quantities $P(j,k)$ and $c(k)$ (or $c_{\mathrm{eff}}(k)$) simultaneously.

\section{Algorithm}
\label{sec:algorithm}

As already stated, there exists an algorithm to create networks with a given degree distribution and a given level of clustering published by \citeauthor{Serrano2005} \cite{Serrano2005}. We incorporated some of their basic ideas into our approach which additionally fixes the degree-degree correlations besides the degree-dependent clustering.

The overall scheme of the algorithm to construct a network with $N$ vertices and a given $P_{d}(j,k)$ and $c_{d}(k)$, $P_{d}(j,k)$ being the number of connections between vertices with degrees $k$ and $j$ (double that number if $k=j$), and $c_{d}(k)$ being the number of triangle edges constituted by vertices with degree $k$, is the following:

We begin by assigning a number of stubs (the target degree) to every vertex according to the degree distribution $P_{d}(k)$, which is calculated from $P_{d}(j,k)$ as $P_{d}(k)=\sum_{j}P_{d}(j,k)/k$.

The next step is to get a list of degrees of triangle-corners, which shall contain $c_{d}(k)$ entries with value $k$.
We also get a copy $P'_{d}$ of $P_{d}(j,k)$ and $c'_{d}$ of $c_{d}(k)$, which are dynamical quantities in the sense that these shall be decreased with every connection and triangle build.
Thus for every connection built we decrease the appropriate entry in the $P'_{d}(j,k)$ matrix by $1$ and for every triangle built (for every connection we place, we check for simultaneous neighbors of the involved vertices as any shared neighbor accounts for a new triangle built) to delete one entry from the triangle list and to decrease $c'_{d}(k)$ by $1$ for every degree involved.

Then we start to build all triangles in the triangle list one by one. Let $v_{i}$ be the vertices involved and $k_{i}$ their target degree.
\begin{enumerate}
\item We draw a random entry $k_{1}$ from the triangle list and draw a corresponding vertex $v_{1}$ with at least one free stub. If we cannot find such a vertex, we delete all entries with value $k_{1}$ from the triangle list and start again.
\item Now, we choose with uniform probability either (a) an edge or (b) a stub of vertex $v_{1}$ out of a list created by omitting all edges for whose end vertex no more triangles can be build (i.e. $c'_{d}(k)=0$). In case of (a), we have chosen an edge and the end vertex is $v_{2}$. If we have drawn a stub (b), we get a vertex $v_{2}$ in the same manner as we got vertex $v_{1}$ with the further condition $P'_{d}(k_{1},k_{2}) > 0$. If it is not possible to find a $k_{2}$ fulfilling this condition, we delete all entries with value $k_{1}$ from the triangle list and start again.
\item Next, we draw (a) an edge or (b) a stub of vertex $v_{2}$ from a list like we did in the preceding step for vertex $v_{1}$, but with edges inserted into the list only if they are fulfilling the supplementary condition of $P'_{d}(k_{1},k_{3})>0$ or vertex $v_{3}$ being connected to vertex $v_{1}$ and vertices $v_{1}$, $v_{2}$, and $v_{3}$ not already constituting a triangle. Having drawn an edge (a), we close the triangle by adding the missing edges and updating all dynamic quantities. Having drawn a stub (b), we choose a $k_{3}$ from the triangle list consistent with $k_{1}$ and $k_{2}$. It might happen that this is not possible and we start again. When we got a $k_{3}$, we draw a vertex $v_{3}$ which either has enough free stubs or is already connected to vertex $v_{1}$ or $v_{2}$, add the missing edges, and update all dynamic quantities.
\end{enumerate}
Note that in steps $2$ and $3$ the case of two or three degrees being the same has to be properly taken into account in order not to build too much triangles or connections, and that self-connections are forbidden.

Those steps are repeated until we cannot build any triangles anymore. This point may be  defined by a maximum number of successive tries that did not result in a triangle being built or until the triangle list is empty.

Afterwards we build the rest of the graph by randomly choosing edges out of the remaining edge list, which contains $P'_{d}(k_{1},k_{2})$ entries ($k_{1}$,$k_{2}$) for all degrees $k_{1}$, $k_{2}$. We choose randomly two non-identical vertices with stubs left and build the edges (if the vertices are not already connected) and delete the edge we chose from the edge list.
We repeat this until the edge list is empty or we cannot find any vertices which still lack connections and are not already connected to each other.
If there are edges we could not build (typically there is no edge left, and very seldomly there are more than one or two edges left), we substitute them by randomly connecting vertices.

\begin{figure}[t]
  \centering
  \includegraphics[width=0.25\textwidth, angle=270]{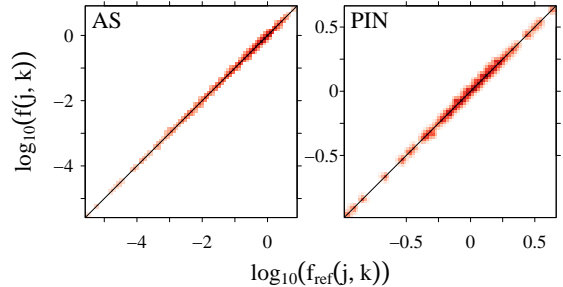}
  \caption{(color online) Plot of the correlation function
    $f(j,k)$ of the random networks generated by the present algorithm vs the correlation
    function $f_{\text{ref}}(j,k)$ of the corresponding empirical network. The data is presented as a density plot. Darker red
    regions contain a higher density of data points, while brighter
    red indicates a lower density.  A reference line $y = x$ is drawn
    as a guide to the eye.}
  \label{fig:corCoeff}
\end{figure}

To validate our algorithm, we use two empirical networks as test
cases: (i) the yeast protein-interaction
network (PIN) constituent of $1,846$ proteins \cite{Jeong2001} downloaded from Barab{\'a}si's web site \mbox{\url{http://www.nd.edu/~networks/}}, (ii) a subset of the internet on the autonomous system level (AS) with $10,515$ vertices (snapshot taken on 03/16/2001) taken from \mbox{\url{http://www.cosin.org/}}.
All self- and multiple-edges were removed from each network.
To test the validity of the algorithm, one measures the joint degree distribution $P(j,k)$ and the degree-dependent clustering $c(k)$ of the empirical networks and uses these functions as input for the construction algorithm. The resulting random
network has to display the same joint
degree distribution $P(j,k)$ (this implies that the degree distribution $P(k)$ is met as well) and the same degree-dependent clustering $c(k)$ as the empirical one. A very sensitive
test to validate if the correlation structure of the reference and the
random network indeed match is on the level of the correlation function $f(j,k)$, which varies on a much smaller scale than the joint degree distribution $P(j,k)$. Thus, comparing the reference
correlation function $f_{\text{ref}}(j,k)$  with the resulting correlation function $f(j,k)$ by use of a correlation coefficient ($1$ means total agreement, $-1$ indicates that the two functions are of opposite sign and $0$ means no
correlation among the two functions in comparison) reveals almost complete agreement of (i) $0.9999(9)$ and (ii) $0.999(7)$. A density plot of the reference function versus the resulting correlation function in Fig.~\ref{fig:corCoeff} verifies the
excellent agreement of the correlation functions $f(j,k)$ and $f_{\text{ref}}(j,k)$, as the density of points is almost solely centered at the diagonal. The statistics per curve are
$10^3$ realizations for the AS network and $10^4$ for the PIN.

\begin{figure}[t]
\centering
\includegraphics[width=0.35\textwidth, angle=270]{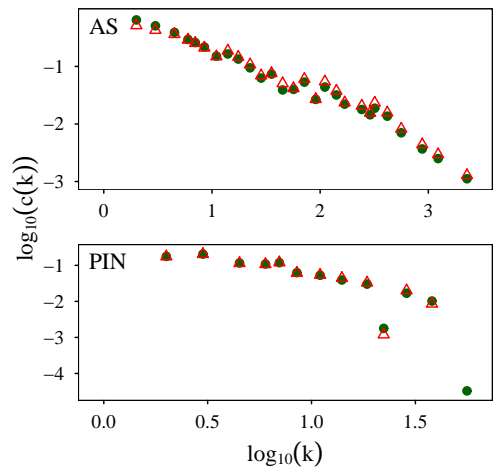} 
\caption{(color online) Degree-dependent clustering coefficient $c(k)$ vs $k$ of empirical graphs (open, red triangles) compared to their randomized versions generated by the present algorithm (full, green circles).}
\label{fig:ckTest}
\end{figure}

However, the main and new point of our algorithm is its ability to conserve the degree-dependent clustering as well.
The quality of agreement is shown in Fig.~\ref{fig:ckTest}. We show a comparison between the degree-dependent clustering $c(k)$ of empirical and generated networks.
One can see that the level of clustering in the PIN and AS network is well reproduced.

\section{Correlations and Clustering}
\label{sec:clusNets}

We wish not only to be able to reproduce correlations and clustering in empirical networks, but also to create graphs from scratch that follow an adjustable correlation pattern expressed by the average nearest neighbor function $k_{\mathrm{nn}}(k)$ and a tunable degree-dependent clustering coefficient $c(k)$. Eq.~\eqref{eq:ckbylambda} defines $c_{\mathrm{eff}}(k)$ as an effective clustering. So we might consider a graph showing
\begin{equation}
\label{eq:constClus}
c_{\mathrm{eff}}(k) = \mu \, ,
\end{equation}
with $\mu$ being a constant between $0$ and $1$, as an equally clustered graph throughout all degree classes. Therefore we may tune the level of clustering by changing $\mu$.
As we are able to control degree-degree correlations by use of the algorithm presented in \cite{Weber2007}, we can calculate the upper limit $\lambda(k)$ and therefore the degree-dependent clustering $c(k)$ from $P(j,k)$ and the target clustering $c_{\mathrm{eff}}(k)$ via Eqs.~\eqref{eq:lambdaDef} and \eqref{eq:ckbylambda}. To get a discrete correlation function $P_{d}(j,k)$, we first create a graph with a given degree distribution $P(k)$ and a correlation structure characterized by a given $k_{\mathrm{nn}}(k)$ using the method presented in \cite{Weber2007}, and obtain its discrete correlation function $P_{d}(j,k)$. With Eq.~\eqref{eq:ckbylambda} we get $c(k)$ and therefore the number of triangles per degree $k$ as
\begin{equation}
 \label{eq:numTri}
c_{d}(k) = c(k)P(k)(k-1)N\, .
\end{equation}
With $c_{d}(k)$, $P_{d}(j,k)$ and the resulting discretisized $P_{d}(k)$ we have the input needed for our algorithm.
To validate the algorithm we tested it for a scale-free graph ($P(k)\propto k^{-\gamma}$) with several levels of clustering and several degrees of assortativity using $k_{\mathrm{nn}}(k) \propto \exp \left((\ln(1 + \frac{k}{k_{\min}}))^{\alpha}\right)$ as an example. The graph size has been set to $N=10^{5}$ vertices. In order to avoid intrinsic degree-degree correlations caused by the constraint of no self- and multiple connections \cite{Catanzaro2005,S.N.2005}, one has to limit the maximum degree to a $k_{\text{max}}$ depending on the scale-free exponent $\gamma$ and the level of (dis-)\-assortativity controlled by $\alpha$ \cite{Weber2007}.

\begin{figure}
\includegraphics[angle=270]{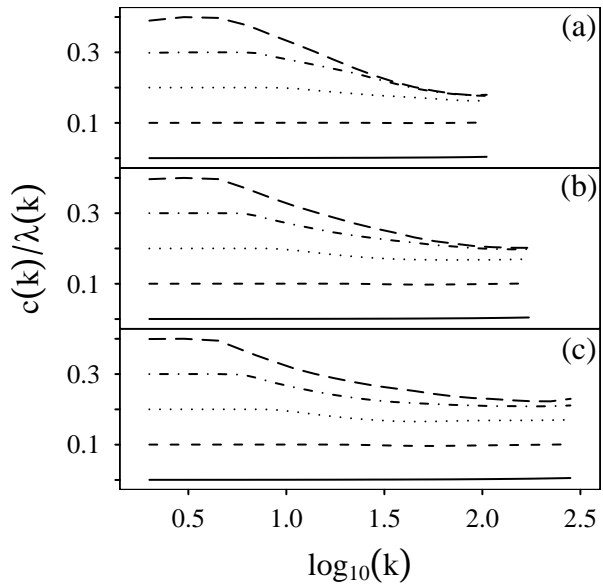}
\caption{Clustering $c_{\mathrm{eff}}(k) = c(k)/\lambda(k)$ vs $k$ examplified for $N=10^{5}$, $\gamma = 2.8$, $k_{\mathrm{nn}}(k)$ as given in the main text, and $\mu  = 0$, $0.1$, $0.2$, $0.3$, and $0.4$ (from bottom to top), (a) $\alpha = 0.2$ (assortative), (b) $\alpha=0$, and (c) $\alpha = -0.2$ (disassortative).}
 \label{fig:ckblk}
\end{figure}

In Fig.~\ref{fig:ckblk} we show the resulting $c_{\mathrm{eff}}(k)$. The statistics per curve are $100$ realizations each, with $P_{d}(j,k)$ drawn for each realization seperately. One observes that a level of clustering close to $\mu=1$ is not achievable with our algorithm. Low levels of clustering are very well reproducible, and medium levels of clustering are very well reproducible for lower degrees, but the higher the degree the more difficult it gets to cross a certain level of clustering, this level being dependent of the level of clustering of the lower degree classes.
This behavior is not surprising as in calculating the upper bound $\lambda(k)$ it is assumed that all vertices $i$ with a degree $k'$ smaller than $k$ have a clustering coefficient $c_{i}=1$. Thus restrictions on the level of clustering of low degree vertices imply stronger restrictions on the level of clustering of high degree vertices. As changing the assortativity via $\alpha$ has only a minor effect on the effective clustering $c_{\mathrm{eff}}(k)$ which can be reached, it seems that the effects of degree-degree correlations on clustering are well described by the upper bound $\lambda(k)$.

\section{Conclusion}
\label{sec:conclusion}

In summary, we have presented an algorithm which generates networks with an {\em a priori} fixed degree-degree correlation structure defined by the joint degree distribution $P(j,k)$ and an adjustable level of clustering defined by the degree-dependent clustering coefficient $c(k)$. As clustering and degree-degree correlations are suspected to play an important role in many dynamical processes taking place on networks, our algorithm may provide a very useful tool to systematically research the influences of those topological properties on different dynamics.

\bibliography{gen_clusNets}

\end{document}